\documentclass[psfig]{aa}
\usepackage{times}
\usepackage{psfig}

\newcommand{\bd}{\begin{displaymath}}
\newcommand{\ed}{\end{displaymath}}
\newcommand{\be}{\begin{equation}}
\newcommand{\ee}{\end{equation}}

\begin{document}

\title{Instability of an accretion disk with a magnetically driven wind}

\author{Xinwu Cao\inst{1,2},\and H.C.\ Spruit\inst{2}}

\offprints{H.C. Spruit(henk@mpa-garching.mpg.de)}

\institute{Shanghai Astronomical Observatory, Chinese Academy of
           Sciences, 80 Nandan Road, Shanghai, 200030, China, 
           and National Astronomical Observatories, 
           Chinese Academy of Sciences, China \\
\and
           Max-Planck-Institut f\"ur Astrophysik,
           Postfach 1317, 85741 
           Garching bei M\"unchen, Germany}
           
\date{Received ; accepted }

\markboth
{Instability of an accretion disk with a magnetically driven wind}
{Instability of disk with wind}

\abstract{
We present a linear analysis of the stability of accretion 
disks in which angular momentum is removed by the magnetic torque
exerted by a centrifugally driven wind. The effects of the dependence
of the wind torque on field strength and inclination, the
sub-Keplerian rotation due to magnetic forces, and the compression of 
the disk by the field are included. A WKB dispersion relation 
is derived for the stability problem. We find that the disk is
always unstable if the wind torque is strong. At lower wind torques
instability also occurs provided the rotation is close to Keplerian. The  
growth time scale of the instability can be as short as the orbital
time scale. The instability is mainly the result of the
sensitivity of the mass flux to changes in the inclination of the field 
at the disk surface. Magnetic diffusion in the disk stabilizes if the 
wind torque is small.
}
\maketitle

\section{Introduction}
        
Magnetically driven winds from accretion disks have been considered 
as promising explanations for highly collimated jets observed in many 
classes of astrophysical objects, ranging from active galactic nuclei 
(AGNs) to young stellar objects (YSOs)(see the reviews by 
Blandford 1993, 2000; Pringle 1993; Spruit 1996). Hydromagnetic winds from
accretion disks provide a mechanism to produce collimated jets as well
as an efficient angular momentum loss mechanism from accretion 
disks.  The jets are powered by the gravitational energy released by 
accretion of matter through a magnetic field with an open configuration. 
If the inclination angle of the magnetic field lines is sufficient 
large with respect to the axis of the accretion disk, a magnetically 
controlled, centrifugally accelerated wind is naturally driven from
the disk surface. Open magnetic field lines threading the accretion disks 
are a crucial ingredient in this model (Blandford \& Payne 1982).
A global, ordered magnetic field is usually assumed in investigations,
though the origin of the ordered magnetic field that is assumed to
thread the disk is still not clear. One possibility is that  
the field is advected inwards by the accreting matter in the disk 
(Blandford \& Payne 1982). The other one is that the field is   
generated locally by a dynamo (e.g. Tout \& Pringle 1996). 
A steady open magnetic field may be maintained if the inward dragging 
of field lines by the disk is balanced everywhere by the outward
motion of field lines due to magnetic diffusivity (Lubow, Papaloizou, 
\& Pringle 1994a).  Lubow, Papaloizou, \& Pringle (1994b, hereafter
LPP) explored the stability properties of the disk-wind 
system on the assumption that the angular momentum of accretion disk
is  removed by the magnetic torque alone. They argue that such a 
disk wind is unstable. K\"onigl \& Wardle (1996) on the other hand
claim stability of their own disk-wind solutions (Wardle \& K\"onigl 
1993). 
Krasnopolski et al. (1999) also claim stability on the basis
of numerical simulations. Instability of the type proposed by LPP has been
incorporated into a numerical model by Agapitou and Papaloizou (in preparation).

\section{Model}

Magnetically driven winds can remove angular momentum from the disk. 
The wind acceleration is sensitive to the inclination of magnetic 
field at the disk surface (Blandford \& Payne 1982). If the magnetic 
field is advected inwards by the accreted matter, the field 
inclination at the disk surface is mainly determined by radial 
velocity of the matter in the disk. A small perturbation increasing the
radial velocity causes the poloidal field to be bent closer to the
disk surface, so that it gives rise to a higher mass loss. By the 
increased angular momentum loss, the radial velocity continues to increase. 
Thus an instability arises (LPP). 

A number of factors have to be included in the analysis. If the
magnetic field strength increases inward in the disk, it exerts
an outward radial force providing partial support against gravity, 
so that the rotation of accretion flow is sub-Keplerian. The height 
of the disk is compressed by both the vertical component of 
gravitational force and the vertical pressure of the radial magnetic 
field component. 

In principle, one can numerically calculate the disk 
structure and the magnetic field configuration in the disk 
simultaneously (Ogilvie 1997, Ogilvie \& Livio 1998). For simplicity, 
in this work we use an approximate
field configuration to estimate the scale-height of an isothermal 
disk compressed by the field.

We assume that the accreting matter is corotating with
magnetic field lines in the disk. This is an appropriate limit,
if the field is sufficiently strong (e.g. Ogilvie \& Livio 2001). 
The matter at the disk surface is centrifugally accelerated 
along the magnetic field line anchored at the disk surface. The matter
corotates with the magnetic field line roughly till it approaches 
Alfv{\'e}n point. Beyond the Alfv{\'e}n radius, the magnetic field exerts no 
further torque on the escaping wind matter. The magnetic torque can be
estimated if the mass loss rate in the wind and the Alfv{\'e}n radius are
known. As usual, the mass loss rate in the wind is determined by the density 
at the slow sonic point in the wind. The slow sonic speed is close to the sound
speed, if the  Alfv{\'e}n radius $r_{\rm A}\gg r_{\rm s}$. To avoid solving 
the equation of radiative transfer in vertical direction of the disk, 
we approximate the temperature of the disk to be a constant along the 
vertical direction. 

For most of the stability analysis, a local approximation can be made 
involving the values of physical quantities near the foot point of a
field line. The analysis, however, involves the magnetic torque in
an essential way, hence the value of the Alfv\'en radius $r_{\rm A}$ for the field
line must be known. Since the Alfv\'en radius depends on the global field
configuration, not only on the field strength at the base of the field line,
it can be regarded as an external parameter for the stability analysis. 
To minimize the number of additional parameters in the analysis, we assume 
the field at the Alfv\'en surface to be roughly self-similar, varying as a 
power of distance from the axis. The value of $r_{\rm A}$ can then be 
determined and we can evaluate the torque exerted by the wind on the
disk. It depends on the mass flux along the field line, which depends on the
density at the sonic point. This in turn depends sensitively on both the
temperature of the disk and on the inclination and strength of the field near
the disk. 

Changes in torque exterted cause the mass distribution in the disk to change.
This in turn causes changes in the distribution of the field lines anchored in
the disk. Since the mass and angular momentum loss rate on a field line 
depends sensitively on the field configuration (in particular the inclination),
there is a strong intrinsic feedback, and it is the purpose of our investigation
to see under which conditions this feedback leads to instability of the type
proposed by LPP. 

The stability problem is in principle global in nature. We localize it 
by a WKB approximation (short wavelength limit) and a dispersion relation 
is then derived for the stability problem. This is analogous to the so-called 
tight-winding approximation used in studies of self-gravitating disks. 

A further simplification we make is that the magnetic torque in a 
steady wind is also be used in the perturbations. This 
assumption is valid if the time scale of the instability is much
longer than the disk-wind coupling time scale. 
We discuss this further in the last section. 

Since our focus is on the instability caused by the wind torque,  
we leave out angular momentum transported by the usual disk viscosity,
so that viscous modes are also absent in stability analyses.   
The radial pressure gradient in the momentum equation is neglected for
similar reasons, to avoid the presence of an acoustic mode. The 
circular motion of the disk in equilibrium is given
by the balance between the gravity and radial magnetic force. 
To evaluate the perturbations  in the magnetic field above the disk we
assume this field to be potential. This a good approximation as long
as the Alfv\'en surface is far from the disk, since the short-wave
approximation ensures that the magnetic perturbations decay quickly
with height above the disk.
The perturbed magnetic field configuration above the disk is then
determined by the normal component of the perturbed field 
at the disk surface. This, in turn, follows from the radial displacements
in the disk, through the induction equation. With respect to the magnetic
field and the forces it exerts, the model is thus the same as that used 
earlier by Tagger et al. (1990), Spruit and Taam (1990), Lubow and Spruit
(1995), Spruit et al. (1995), and Stehle and Spruit (2001).

To evaluate the mass loss rate in the magnetic wind, a model for the distribution 
of mass with distance above the midplane is needed. We assume
the Gaussian distribution valid for an isothermal disk, but also include the
effect on the disk thickness of vertical confinement by the magnetic pressures.

Strictly speaking, the presence of a magnetic torque will lead to an
azimuthal component of the field. However, the azimuthal field component
at the disk surface is always small compared with the radial component
for any wind models as long as the  Alfv{\'e}n radius is far from the radius
of the field footpoint at the disk mid-plane (see further discussion in section 3.2). 
If the magnetic field strength inside the disk is suffiently weak, differential
rotation acting on the $r$- and $z$-components will produce a time-dependent
azimuthal field. Such a nonstationary state will lead to magnetic turbulence 
such as found in numerical simulations of an initially weak field
(Hawley et al. 1995). The field we consider here, on the other hand, is a 
systematic poloidal field (i.e. crossing the disk plane), and sufficiently 
strong such that Balbus-Hawley instability is suppressed. At this strength, the
field is also sufficiently strong to enforce `isorotation': constancy of the 
rotation rate along field lines. Though the origin of such fields is not as
clear as the turbulent fields that evolve from initially weak seed fields,
they are by far the most logical configurations for producing systematic outflows
from accretion disks. Spruit et al. (1995), Stehle (1997) and Stehle and 
Spruit (2001) have shown that even surprisingly strong poloidal fields, approaching 
(a fraction of) equipartition with orbital kinetic energy, can still be stably
anchored in a disk. These analyses did not include instabilities associated
with magnetically driven outflows however, which are the subject of the present 
study.

\section{Magnetic torque}

From standard magnetic wind theory, we know that the
angular momentum flux along a field line is given by
$T=\dot m \Omega(r_{\rm i})r_{\rm A}^{2}$, where $r$ is the
cylindrical radial coordinate, $r_{\rm A}$ the Alfv{\'e}n radius,
$\dot m$ the mass loss rate, and $\Omega$ the rotation rate of the
footpoint at radius $r_{\rm i}$ on the disk. This is the total angular
momentum flux, including both the magnetic torque and the flux carried
by the mass itself. For the effect on the disk, only the magnetic torque
$T_{\rm m}$ is consequential; evaluating it at the disk surface, and
counting both surfaces, we have

\be
T_{\rm m}=2 \dot m \Omega_0(r_{\rm i})(r_{\rm A}^{2}-r_{\rm i}^2),
\ee
In the limit $r_{\rm A}\gg r_{\rm i}$, $T_{\rm m}\sim 
2\dot m \Omega_{0}(r_{\rm i})r_{\rm A}^{2}$.

The mass loss rate in the wind is governed by the position of the sonic
point:

\be
\dot m\sim \left({\frac {B_z}{B_{\rm p}}} \right) 
\rho c_{\rm s}|_{z=z_{\rm s}},
\ee
where $z_{\rm s}$ is the height of the sonic point above the midplane
of the disk.

As discussed in section 2, the mass loss rate of wind from an 
isothermal disk is estimated as 

\be
\dot m \sim  \left({\frac {B_z}{B_{\rm p}}} \right)
\rho_{0}c_{\rm s}\exp[-(\Psi_{\rm es}-\Psi_{\rm ei})/c_{\rm s}^2],
\ee
where $\Psi_{\rm es}$ and $\Psi_{\rm ei}$ are the effective potential 
at the sonic point and the footpoint of the  magnetic field line 
respectively. The effective potential $\Psi_{\rm e}$ is given by 

\be
\Psi_{\rm e}(r, z)=-{\frac {GM}{(r^2+z^2)^{1/2}}}
-{1\over 2}{\Omega^2(r_{\rm i})}r^2,
\ee
where $r_{\rm i}$ is radius of the magnetic field line footpoint 
at mid-plane of the disk. 

\begin{figure}
\centerline{\psfig{figure=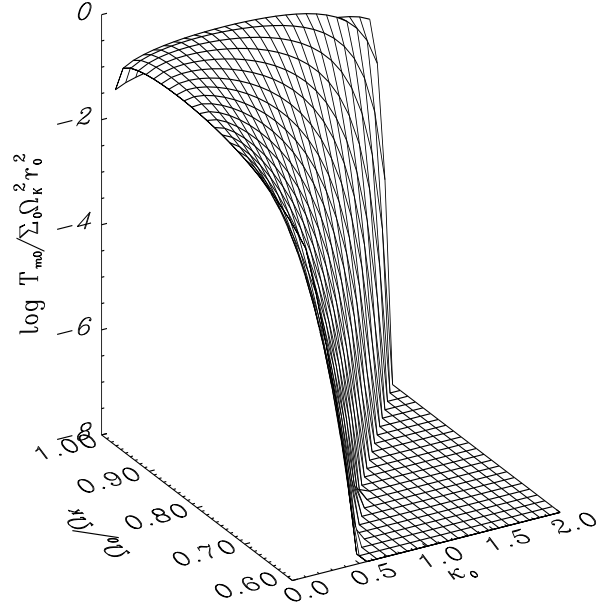,width=10.0cm,height=10.0cm}}
\caption{The magnetic torque as functions of the dimensionless 
angular velocity of the disk and the magnetic inclination 
$\kappa_0$, for a disk thickness  $H/r=0.01$. The angular 
velocity is also a measure of the field strength
($1-(\Omega_0/\Omega_{\rm K})^2\sim B^2$).
}
\end{figure}

\begin{figure}
\centerline{\psfig{figure=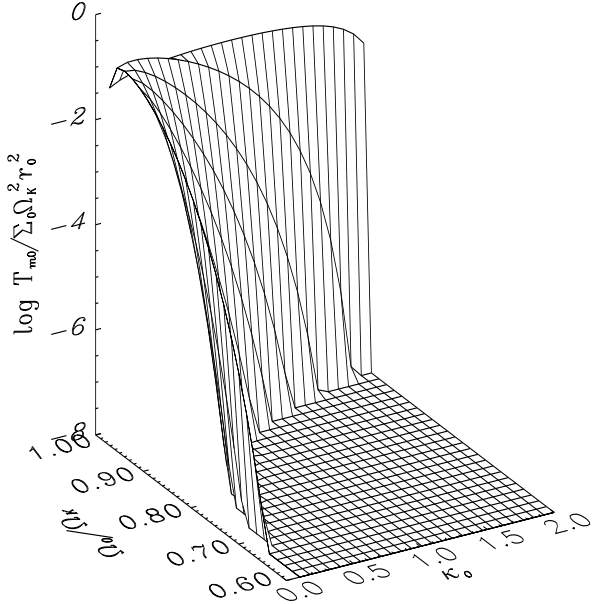,width=10.0cm,height=10.0cm}}
\caption{Same as Fig. 1, but for $H/r=0.001$.
}
\end{figure}

The position of the sonic point can be estimated, if we know the shape
of the magnetic field lines between the footpoint and the sonic point. 
In principle, the field line shape is computable by solving the
radial and vertical momentum equations together with suitable boundary
conditions. Analytic solutions are available only for two 
extreme cases, the weak and strong field extremes. We find that the
field line shapes are slightly different in these two extreme cases. 
In the strong field case, the disk is compressed mainly by the
vertical component of magnetic force, and the shape is similar to 
the Kippenhahn-Schl\"uter model (KS) (Kippenhahn \& Schl\"uter 1957), valid for
an isothermal sheet suspended against gravity by a magnetic field.   
The exact shape of the field lines depends on the thermal structure of the
disk. For our purposes an isothermal layer would be a sufficient model, and
the KS model applicable. Its analytic form is still a
bit too cumbersome however, so we have approximated it by the following
expression:

\be
r-r_{\rm i}={\frac {H}{\kappa_0\eta_{\rm i}^2}}(1-\eta_{\rm
i}^2+\eta_{\rm i}^2 z^2H^{-2})
^{1/2}-{\frac {H}{\kappa_0\eta_{\rm i}^2}}(1-\eta_{\rm i}^2)^{1/2},
\ee
where $H$ is the scale height of the disk, $\eta_{\rm i}=\tanh(1)$. 
The inclination of the field line  $\kappa=B_z/B_r^{\rm S}$ is a function 
of $z$. By the index $_0$ we denote the value $\kappa_0$ measured at the 
nominal disk surface $z=H$. High above the disk, the inclination then 
becomes $\kappa_0\tanh(1)$. We find that this expression reproduces the 
basic features of the KS model well for both weak and strong field cases. 

We approximate the position of the sonic point
as the maximum of the effective potential along the magnetic field line.
The height above the disk of the sonic point at radius $r_0$ of the disk 
is then from using the potential (4) and the magnetic field line shape (5):

\bd
z_{\rm s}\simeq 
{H\over {\eta_{\rm i}} }\left\{\right.
\ed
\be
\left. \left[ {\frac
{\kappa_0\eta_{\rm i}^2r_0(1-\tilde\Omega_0^2)+3\tilde\Omega_0^{2}
H(1-\eta_{\rm i}^2)^{1/2}} {3\tilde \Omega_0^2H
-\kappa_0^2\eta_{\rm i}^2H}}\right ]^2-1+\eta_{\rm i}^2\right\}^{1/2}, 
\ee
which is valid for $z_{\rm s}/r_0\ll 1$.

To complete the estimate, we need the value of the  scale height of 
the disk $H$, which can be calculated from the vertical component 
of momentum equation

\be
c_{\rm s}^2{\frac {d\rho(z)}{dz}}=
-\rho(z)\Omega_{\rm K}^2 z-{\frac {B_r(z)}{4\pi}}
{\frac {dB_r(z)}{dz}}.
\ee
The density distribution $\rho(z)$ in $z$-direction can be
obtained by integrating Eq. (7) with the given field configuration
and the assumed disk temperature.

As a result of the magnetic
forces the distribution is not exactly the Gaussian distribution 
of an isothermal disk.
The disk thickness is reduced by the magnetic forces.
In this case, we define the disk scale height $H$ as
$\rho(H)=\rho(0)\exp(-1/2)$, the disk scale height $H$ can then be
evaluated numerically.
From these numerical values, we have constructed
a fitting formula used in the rest of the calculations
to represent the parameter dependence of the scale height $H$:
\bd
H={1\over 2}\left [ {\frac {4c_{\rm s}^2}{\Omega_{\rm K}^2}}
+{\frac {(\Omega_{\rm K}^{2}-\Omega_0^2)^2r_0^2}{4(1-{\rm e}^{-1/2})^2
\kappa_0^2\Omega_{\rm K}^{4}}}\right]^{1/2}
\ed
\be
-{\frac {(\Omega_{\rm K}^{2}-\Omega_0^2)r_0}{4(1-{\rm e}^{-1/2})\kappa_0
\Omega_{\rm K}^{2}}}.
\ee
We find that the scale height $H$ given by expression (8)
is accurate to about 1\% compared with the numerical results in the
range of the parameters adopted in this paper.  
The actual density distribution deviates from a Gaussian distribution, so
the density at the sonic point still is only approximate, but
the fitting formula properly takes into account the compression
of the disk in the strong field case. 
Note that Eq. (8) reduces to $H=c_{\rm s}/\Omega_{\rm K}$ in the
Keplerian case. Now, using Eqs. (4)-(6), we obtain the effective 
potential difference $\Delta\Psi_{\rm e}$ between the sonic point 
and the magnetic footpoint in the mid-plane of the disk 

\bd
\Delta\Psi_{\rm e}=\Psi_{\rm es}-\Psi_{\rm ei}
={1\over 2}\Omega_{\rm K}^2 z_{\rm s}^2
\ed
\bd
+{\frac {(\Omega_{\rm K}^2-\Omega_0^2)r_0H}
{\kappa_0\eta_{\rm i}^2}}
\left[ (1-\eta_{\rm i}^2+\eta_{\rm i}^2z_{\rm s}^2H^{-2})^{1/2}
-(1-\eta_{\rm i}^2)^{1/2}\right]
\ed
\be
+\left( \Omega_{\rm K}^2-{\frac {5}{2}}\Omega_0^2\right)
{\frac {H^2}{\kappa_0^2\eta_{\rm i}^4}}
\left[ (1-\eta_{\rm i}^2+\eta_{\rm i}^2z_{\rm s}^2H^{-2})^{1/2}
-(1-\eta_{\rm i}^2)^{1/2}\right]^2.
\ee
Using Eqs. (3)-(9), we can evaluate the mass loss rate in wind. 
The magnetic field far from the disk surface 
is assumed to be roughly self-similar as discussed in section 2:

\be
B_{\rm p}^{\rm A}\sim B_{z}(r_0)f\left({\frac
{r_{\rm A}}{r_0}}\right)=B_{z}(r_0)\left
({\frac {r_{\rm A}}{r_0}}\right)
^{-\alpha}, \alpha>1,
\ee
where $B_{z}(r_0)$ is the $z$-component of the magnetic field at the disk
surface.

At the Alfv\'en point, the Alfv{\'e}n velocity is 
\be
v_{\rm A}^{\rm A}={\frac {B_{\rm p}^{\rm A}}{(4\pi\rho_{\rm A})^{1/2}}},
\ee
As in all magnetic acceleration models, its value is of the order
$v_{\rm A}^{\rm A}\sim r_{\rm A}\Omega_{0}$,
where $B_{\rm p}^{\rm A}$ and $\rho_{\rm A}$ are the poloidal magnetic
field strength and density of the outflow at Alfv{\'e}n point. 

Mass and magnetic flux conservation along a magnetic field line
requires

\be
{\frac {\dot m}{B_{z}^{\rm d}}}\sim {\frac {\rho_{\rm A}v_{\rm
A}^{\rm A}}{B_{\rm p}^{\rm A}}}.
\ee

Combining Eqs. (10)-(12), the Alfv{\'e}n radius is found: 

\be
r_{\rm A}\simeq \left[ {\frac { {B_{z}^{\rm d}}^2} {4\pi\dot m
\Omega_{0}(r_0)}}
\right]^
{\frac {1}{\alpha+1}}r_0^{\frac {\alpha}{\alpha+1}}.
\ee

Now we can evaluate the magnetic torque $T_{\rm m}$ by using Eqs (3), (6)
and (13). 

We define dimensionless quantities by 
\be
{\tilde T}_{\rm m}={ {T_{\rm m}}\over{\Sigma_0\Omega_{\rm K}^{2} r_0^2}},\qquad
\tilde \Omega_0={{\Omega_0}\over {\Omega_{\rm K}}},\label{ttm}\qquad
\tilde H={H\over r_0}.\label{diml}
\ee
The dimensionless magnetic torque $\tilde T_{\rm m}$ is then given by 
\bd
{\tilde T}_{\rm m}=
\tilde \Omega_0^{\frac {\alpha-1}{\alpha+1}}
(1-\tilde\Omega_0^2)^{\frac {2}{\alpha+1}}
\kappa_0^{\frac {2}{\alpha+1}}
\kappa_{\rm s}^{\frac {\alpha-1}{\alpha+1}}
(1+\kappa_{\rm s}^2)^{-{\frac {\alpha-1}{2(\alpha+1)}}}
\ed
\be
\left [1+{\frac {1-\tilde\Omega_0^2}
{2(1-{\rm e}^{-1/2})\kappa_0\tilde H}}\right ]
^{\frac {\alpha-1}{2(\alpha+1)} }
\exp\left(-{\frac {\alpha-1}{\alpha+1} }
{\frac {\Delta \Psi_{\rm e}}{c_{\rm s}^2}}\right), 
\ee
The parameters of the model are now $\tilde H$, $\tilde\Omega_0$, 
$\kappa_0$ and the self-similar index $\alpha$ of magnetic field shape.
Here $\kappa_{\rm s}$ is the inclination of the field line at the 
sonic point.  

\subsection{Dependence of the magnetic torque on field strength 
and inclination}

With increasing field strength, the rotation of the disk becomes 
more sub-Keplerian. Though the increasing field strength itself
increases the torque through the increasing  Alfv{\'e}n radius 
[second factor in Eq. (15)], the lower rotation of the footpoints 
also increases the potential barrier that the wind has to overcome 
(last factor). In many cases of interest, an increasing field strength
will reduce the magnetic torque (e.g. Shu 1991; Ogilvie \& Livio 1998,
2001). In the following, we use the rotation rate relative to Keplerian,
$\tilde\Omega_0$, as measure of the field strength. Figures 1 and 2 
show expression (15) for the cases $\tilde H=0.01$ and $\tilde
H=10^{-3}$. 

\subsection{Azimuthal magnetic field component at the disk surface}

The magnetic torque $T_{\rm m}$ exerted on the unit surface
area of the disk is  given by

\be
T_{\rm m}={\frac {B_z B_{\phi}^{\rm S}r}{2\pi}},
\ee
where $B_{\phi}^{\rm S}$ is the azimuthal field component at the
disk surface. Let $g_{\rm m}$ be the radial acceleration due to magnetic tension,
averaged over the surface mass density of the disk,
$g_{\rm m}=B_{r}^{\rm S}B_z/(2\pi\Sigma)$. The radial
balance of forces is then $r\Omega_{0}^{2}=g-g_{\rm m}$. Hence we have the
relation
\be
{\frac {B_{\phi}^{\rm S}}{B_{r}^{\rm S}}}
={\frac {\tilde T_{\rm m0}}{1-\tilde\Omega_0^2}}.
\ee
For the parameter values where our analysis will be valid, the azimuthal field 
strength at the disk surface is small, $B_{\phi}^{\rm S}\ll B_{r}^{\rm S}$  
(see Figs. 1 and 2). 

\section{Disk equations}

As discussed in section 2, we neglect the usual disk viscosity 
in the problem. Torques due to the magnetic wind are therefore the only
source of angular momentum loss of the disk material, and in their absence
the mass flux through the disk vanishes. The mass conservation and angular 
momentum equations for the disk are then 

\be
\left({\partial\over {\partial t}}+{v_r}{\partial\over {\partial r}}\right)
\Sigma+{\Sigma}{{\partial v_r}\over{\partial r}}=0,
\ee
and

 \be
\left({\partial\over {\partial t}}+{v_r}{\partial\over {\partial r}}\right)
v_{\phi}+{{v_{r}v_{\phi}}\over r}
+{{T_{\rm m}}\over {\Sigma r}}=0,
\ee
where $T_{\rm m}$ is the torque contributed by the magnetically driven wind 
per unit area of disk surface. The radial component of the momentum
equation is
\be
\left({\partial\over {\partial t}}+{v_r}{\partial\over {\partial r}}\right)
v_{r}+g-{{v_{\phi}^2}\over r}-g_{\rm m}=0.
\ee
where the gravitational and radial magnetic forces are given 
by $g=GM/r^{2}$ and  $g_{\rm m}=B_{r}^{\rm S}B_z/(2\pi\Sigma)$, and  
we neglect the gas pressure gradient as discussed in section 2.
The magnetic induction equation is 

\be
\left({\partial\over {\partial t}}+{v_r}{\partial\over {\partial r}}\right)
B_{z}+{\frac {v_r B_z}{r}}
+{B_z}{{\partial v_r}\over {\partial r}}
+{\frac {\eta}{r}}{\frac {\partial}{\partial r}}
\left(r {\frac {\partial B_r}{\partial z}}\right)=0.
\ee
where $\eta$ is magnetic diffusivity, $H$ is the disk scale-height. 

As discussed in section 2, a potential field above the disk is
assumed, so that the stream function of the field satisfies 

\be
{{\partial^2\Phi}\over {\partial r^2}}-{1\over r}
{{\partial\Phi}\over {\partial r}}
+{{\partial^2\Phi}\over {\partial z^2}}=0.
\ee

\subsection{Linearized equation in a local approximation}

As discussed in section 2, the problem can be localized by 
a WKB approximation in a short wavelength limit. 
The perturbed stream function is still given by Eq. (22):
\be
{{\partial^{2}\delta\Phi}\over {\partial r^{2}}}
-{1\over r}{{\partial\delta\Phi}\over {\partial r}}
+{{\partial^{2}\delta\Phi}\over {\partial z^{2}}}=0.
\ee
The strength $B_z(r)$ at the disk surface acts as boundary 
condition for this potential problem. Its solution is in general 
a global one, the solution at any point $(r,~z)$ depends on the 
entire field strength distribution $B_z(r)$. In a short wavelength 
limit, however, the problem becomes local again.  Our short wavelength
limit is the magnetic equivalent of the so-called tight-winding limit 
in self-gravitating disks. 

Equation (23) has separable solutions $\delta\Phi(r,z)=F(r)\exp(-k|z|)$,
where $F(r)$ satisfies

\be
F^{\prime\prime}-{1\over r}F^{\prime}+k^{2}F=0.
\ee
Let $y=k(r-r_{0})$. In the short-wave limit $kr\gg 1$, we have

\be
{{d^{2}F}\over {dy^{2}}}-\epsilon{{dF}\over {dy}}+F=0,
\ee
where $\epsilon=1/kr$ ($\epsilon\ll 1$). 

Denote Eulerian perturbations of a quantity $q$ by $\delta q$, Lagrangian
perturbations by an index $_1$, and the equilibrium state by an index $_0$.
 
The perturbed fraction of the stream function is: 

\be
\delta\Phi(r,z)=A\exp\left({{\epsilon y}\over 2}\pm iy-k|z|\right).
\ee
The perturbed components of the magnetic field are 

\be
\delta B_{r}(r, z)=-{1\over r}{{\partial \delta\Phi}\over {\partial z}}
={k\over r}\delta\Phi(r, z),
\ee
and

\be
\delta B_{z}(r, z)={1\over r}{{\partial \delta\Phi}\over {\partial r}}
={1\over r}\left({{\epsilon k}\over 2}+ ik\right)\delta\Phi (r, z).
\ee
Here we have taken the sign in Eq. (26) to be ``$+$''. 
The magnetic field components are related by 

\be 
\delta B_{z}=\delta B_{r}\left({\epsilon\over2}+i\right)\approx
i\delta B_{r}.
\ee

Let a dot be the Lagrangian time derivative, and $\xi$ be the Lagrangian 
displacement. Neglecting the gradient of $r$ in the background and assuming
$H$ to be a constant along $r$,  the linearized
equations of the disk are:
\be
\Sigma_{1}+{{\Sigma_0}}\partial_{r}\xi_r=0,
\ee

\be
\ddot{\xi}_{\phi}+(2\Omega_{0}+S)\dot\xi_{r}
+{{T_{\rm m1}}\over {\Sigma_{0}r}}
-{{T_{\rm m0}}\over{\Sigma_{0}^{2}r}}\Sigma_{1}=0
\ee
where the shear rate $S=r (d\Omega_0/dr)$.

\be
\ddot{\xi}_{r}-2\Omega_{0}\dot{\xi}_{\phi}-g_{\rm m1}=0,
\ee
where  $r\Omega_{0}^{2}=g-g_{\rm m}$,

\be
{\dot B}_{z1}+B_{z0}\partial_{r}{\dot\xi}_{r}
+{\frac {\eta k} {H}} B_{z1}=0.
\ee

As discussed in section 2, we assume that the magnetic torque as evaluated 
for steady conditions are also valid for the perturbations. The magnetic torque 
$T_{\rm m}$ is a function of the angular velocity of the flow $\Omega$ and 
the magnetic field inclination 
$\kappa_0$, $\kappa_{0}=B_{z}/B_{r}^{\rm S}$, at the disk surface. The perturbed 
fraction of the magnetic torque $\delta T_{\rm m}$ is given by

\begin{eqnarray*}
\delta T_{\rm m}&=& {\frac {\partial T_{\rm m}(\kappa, \Omega)}{\partial\kappa}}
 \delta\kappa+{\frac {\partial T_{\rm m}(\kappa, \Omega)}{\partial\Omega}}
\delta\Omega
\\
&=& {\frac {\partial T_{\rm m}(\kappa, \Omega)}{\partial \kappa}} \left(
{{\delta B_{z}}\over {B_{r0}}}+i{{B_{z0}\delta B_{z}}\over {B_{r0}^2}} \right)
+{\frac {\partial T_{\rm m}(\kappa,
\Omega)}{\partial\Omega}}\delta\Omega
\\
&=& {\frac {\partial T_{\rm m}(\kappa, \Omega)}{\partial \kappa}}
(\kappa_{0}+i\kappa_{0}^2){{\delta B_{z}}\over {B_{z0}}}
+{\frac {\partial T_{\rm m}(\kappa,
\Omega)}{\partial\Omega}}\delta\Omega.
\end{eqnarray*}
where Eq. (29) is used. The perturbed fraction  of the radial magnetic
force is given by 

\be
\delta g_{\rm m}= g_{\rm m0}\left( {\frac {\delta B_{r}}{B_{r0}}}+{\frac
{\delta B_{z}}{B_{z0}}}-{\frac {\delta \Sigma}{\Sigma_0}}\right ).
\ee

Neglecting the radial gradients of the equilibrium quantities (a local 
approximation) the Eulerian and Lagrangian variations are equivalent, so that
the perturbed torque and magnetic acceleration are
\be
T_{\rm m1}= {\frac {\partial T_{\rm m}(\kappa, \Omega)}{\partial \kappa}}
(\kappa_{0}+i\kappa_{0}^2){{B_{z1}}\over {B_{z0}}}
+{\frac {\partial T_{\rm m}(\kappa,
\Omega)}{\partial\Omega}} { {\dot\xi_{\phi}}\over {r_0}}
\ee
and
\be
g_{\rm m1}= g_{\rm m0}\left[(1-i\kappa_0){{B_{z1}}\over {B_{z0}}}
+\partial_r\xi_r\right],
\ee
where Eq. (30) has been substituted.
Combining Eqs. (30)-(33), we have

\bd
{\partial_{tt}\ddot\xi}_{r}
+\left({1\over {\Sigma_{0}r^2}}{\frac {\partial T_{\rm m}}{\partial \Omega}}
+{\frac {\eta k}{H}}\right)\partial_t\ddot\xi_{r}
-i\kappa_{0}g_{\rm m0}\partial_{r}\ddot\xi_{r}
\ed
\bd
+\left[2\Omega_{0}(2\Omega_{0}+S)
+{\frac {\eta k} {\Sigma_0r^2H} } {\frac {\partial T_{\rm m}}{\partial\Omega}}
\right]\ddot\xi_{r}
\ed
\bd
-\left[{ {2\Omega_{0}}\over{\Sigma_{0}r}}
{\frac {\partial T_{\rm m}}{\partial \kappa}}
(\kappa_0+i\kappa_0^2)
-{{2\Omega_0T_{\rm m0}}\over {\Sigma_{0}r}}
+i{\frac {g_{\rm m0}\kappa_0}{\Sigma_0r^2}}
{\frac {\partial T_{\rm m}}{\partial\Omega}}\right.
\ed
\bd
\left.
+{\frac {\eta k g_{\rm m0}} {H}}\right]\partial_r\dot\xi_r
+{\frac {2\eta k\Omega_0} {H}}(2\Omega_0+S)\dot\xi_r
\ed
\be
-\left({\frac {\eta k g_{\rm m0}}{\Sigma_0 r^2 H}}
{\frac {\partial T_{\rm m} } {\partial \Omega}}
-{\frac {2\eta k \Omega_0 T_{\rm m0}}{\Sigma_0 rH}}\right)
\partial_r\xi_r=0.
\ee

\section{Dispersion relation}

We write the perturbed quantities as 
\bd
\xi_{r} \sim \exp[i(\omega t+kr)].
\ed
The dispersion relation follows from Eq. (37):

\bd
\omega^4-i\left( {\frac {1}{\Sigma_{0}r^2}}
{\frac {\partial T_{\rm m}}{\partial \Omega}}
+{\frac {\eta k}{H}}\right)\omega^3
\ed
\bd
-\left[\kappa_{0}g_{\rm m0}k+2\Omega_{0}(2\Omega_{0}+S) 
+{\frac {\eta k}{\Sigma_0r^2H}} {\frac {\partial T_{\rm m}}{\partial \Omega}}
\right]\omega^2
\ed
\bd
+\left[ {{2k\Omega_{0}}\over {\Sigma_{0}r}}
{\frac {\partial T_{\rm m}}{\partial \kappa}}
(\kappa_{0}+i\kappa_{0}^2)
+i{\frac {g_{\rm m0} \kappa_0 k}{\Sigma_{0}r^2}}
{\frac {\partial T_{\rm m}}{\partial \Omega}}\right.
\ed
\bd
\left.
-{2k\Omega_{0}{T_{\rm m0}}\over {\Sigma_{0}r}}
+{\frac {\eta k^2 g_{m0}}{H}}
+i{\frac {2\eta k\Omega_0}{H}}(2\Omega_0
+S)\right]\omega
\ed
\be
-i{\frac {\eta k^2 g_{\rm m0}}{\Sigma_0r^2H}}
{\frac {\partial T_m}{\partial \Omega}}
+i{\frac {2\eta k^2 \Omega_0 T_{\rm m0}}{\Sigma_0 rH}}=0.
\ee
We use some dimensionless quantities in addition to those defined 
in Eq. (14):
\bd
\tilde\lambda={\lambda\over H},\qquad\tilde\omega={\omega\over \Omega_{\rm K}},
\qquad\tilde S={{r_0} \over {\Omega_{\rm K}}} {d\Omega_0\over dr},
\ed
\be
\tilde g_{\rm m0}={g_{\rm m0}\over r_0\Omega_{\rm K}^2},\qquad
\tilde\eta ={\eta\over H^2\Omega_{\rm K}},\label{scal}
\ee
and the quantities $p$, $q$: 
\bd
{ {\frac  {\partial \tilde T_{\rm m}} {\partial \tilde \Omega_0} }
={\frac {p}{\tilde \Omega_0}} \tilde T_{\rm m0}  },\qquad
{ {\frac  {\partial \tilde T_{\rm m}}{\partial \kappa}}
={\frac {q}{\kappa_0}}\tilde T_{\rm m0} },
\ed
which measure the dependence of the wind torque on the rotation rate 
and field inclination, respectively. 

Equation (38) becomes
\bd
{\tilde\omega}^{4}-i\left({p\over {\tilde\Omega_0}}\tilde T_{\rm m0}
+{\frac {2\pi\tilde\eta}{\tilde\lambda}}
\right){\tilde\omega}^{3}
-\left[{\frac {2\pi\kappa_0} {\tilde\lambda\tilde H}}
(1-\tilde\Omega_0^2)\right.
\ed
\bd
\left.
+2\tilde\Omega_0(2\tilde\Omega_0+\tilde S)
+{\frac {2\pi p \tilde\eta\tilde T_{\rm m0}}{\tilde\Omega_0}}\right]
\tilde\omega^2
-\left[{\frac {4\pi\tilde\Omega_0\tilde T_{\rm m0}}{\tilde\lambda\tilde H}}\right.
\ed
\bd
-i{\frac {2\pi p\kappa_0\tilde T_{\rm m0}}{\tilde\lambda\tilde\Omega_0
\tilde H}}(1-\tilde\Omega_0^2)
-{\frac {4\pi q\tilde\Omega_0}{\tilde\lambda\tilde H}}
\tilde T_{\rm m0}(1+i\kappa_0)
\ed
\bd
\left. 
-{\frac {4\pi^2\tilde\eta}{\tilde\lambda^2\tilde H}}
(1-\tilde\Omega_0^2)
-i{\frac {4\pi\tilde\eta \tilde\Omega_0}{\tilde\lambda}}
(2\tilde\Omega_0+\tilde S)\right]\tilde\omega
\ed
\be
-i{\frac {4\pi^2\tilde\eta p \tilde T_{\rm m0}}
{\tilde\lambda^2\tilde\Omega_0\tilde H}} (1-\tilde\Omega_0^2)
+i{\frac {8\pi^2\tilde\eta\tilde\Omega_0\tilde T_{\rm m0}}
{\tilde\lambda^2\tilde H}}=0. 
\ee

\subsection{Restrictions on parameter values}

The dispersion relation contains several small parameters, since 
a number of conditions have to be satisfied for consistence with the
assumptions made. For a local approximation to be valid, we must have
\be \lambda/r\ll 1.\ee
On the other hand, since we have approximated the disk as thin, the 
wavelength has to be large compared with the scale height:
\be H \ll \lambda. \ee

An unknown in the problem is the magnetic diffusivity, hence we must explore
the dependence of the results on $\eta$. In our assumptions, we have ignored
viscosity in the disk. If the magnetic diffusivity is due to the same small
scale process that causes the viscosity, it is then most consistent to set 
$\eta=0$ as well. On the other hand, it is instructive to explore the 
consequences of allowing some diffusion, since it is likely to damp
instabilities. With our scalings (39), the relevant value is then
$\tilde\eta\sim\alpha\sim 1$, where $\alpha$ is the viscosity parameter.
We report results for both cases, $\tilde\eta\ll 1$ and $\tilde\eta\sim 1$.

\subsection{Approximate analytic results}

The nature of the unstable modes can be explored by looking 
at limiting cases.  We can look at the 
case  where the magnetic torque is weak, $\tilde T_{\rm m0}\ll 1$. 
Assuming $1/ \tilde \lambda$ and $\tilde T_{\rm m0}$ to be the same order 
and ignoring quadratic and higher terms, the last two terms 
in Eq. (40) containing $\tilde 
T_{\rm m0}/\tilde\lambda$ can be omitted. The dispersion relation reduces to 
a third-order equation. Let $\tilde\omega=\tilde \omega_{\rm r}
+i\tilde\omega_{\rm i}$,  it can then 
be separated into two equations:

\bd
\tilde\omega_{\rm r}^3
-\left[3\tilde\omega_{\rm i}^2
-2\tilde\omega_{\rm i}
\left({\frac {p\tilde T_{\rm m0}}{\tilde\Omega_0}}
+{\frac {2\pi\tilde\eta}{\tilde\lambda}}\right)
+{\frac {2\pi p\tilde\eta\tilde T_{\rm m0}}
{\tilde\Omega_0}}
\right.
\ed
\bd
\left.
+2\tilde\Omega_0(2\tilde\Omega_0+\tilde S)
+{{2\pi\kappa_0}\over{\tilde\lambda\tilde H}}(1-\tilde\Omega_0^2)\right]
\tilde\omega_{\rm r}
+{\frac {4\pi q\tilde\Omega_0} {\tilde\lambda\tilde H}}\tilde T_{\rm m0}
\ed
\be
-{\frac {4\pi \tilde\Omega_0} {\tilde\lambda\tilde H}}\tilde T_{\rm m0}
+{\frac {4\pi^2\tilde\eta}{\tilde\lambda^2\tilde H}}
(1-\tilde\Omega^2_0)=0
\ee
and
\bd
\tilde\omega_{\rm i}^3
-\left({\frac {p\tilde T_{\rm m0}}{\tilde\Omega_0}}
+{\frac {2\pi\tilde\eta}{\tilde\lambda}}\right)
\tilde\omega_{\rm i}^2
\ed
\bd
-\left[ 3\tilde\omega_{\rm r}^2
-2\tilde\Omega_0(2\tilde\Omega_0+\tilde S)
-{{2\pi\kappa_0}\over{\tilde\lambda\tilde H}}(1-\tilde\Omega_0^2)
-{\frac {2\pi p\tilde\eta\tilde T_{\rm m0}}
{\tilde\Omega_0}}\right]\tilde\omega_{\rm i}
\ed
\bd
+\left({\frac {p\tilde T_{\rm m0}}{\tilde\Omega_0}}
+{\frac {2\pi\tilde\eta}{\tilde\lambda}}\right)
\tilde\omega_{\rm r}^2
-{\frac {4\pi q\kappa_0\tilde\Omega_0}{\tilde\lambda\tilde H}}\tilde T_{\rm
m0}
\ed
\be
-{\frac {2\pi p \kappa_0}{\tilde \lambda\tilde H\tilde\Omega_0}}
(1-\tilde\Omega_0^2)\tilde T_{\rm m0}
-{\frac {4\pi\tilde\eta\tilde \Omega_0  }{\tilde\lambda}}
(2\tilde\Omega_0+\tilde S)=0.
\ee

\subsubsection{$\tilde\eta\ne 0$, $\tilde T_{\rm m0}=0$}

We now analyze the marginal stability, i.e., ${\tilde\omega}_{\rm
i}=0$. In the absence of the magnetic torque ($\tilde T_{\rm m0}=0$),  
Eqs. (43) and (44) reduce to

\bd
\tilde\omega_{\rm r}^3
-\left[3\tilde\omega_{\rm i}^2
-{\frac {4\pi\tilde\eta}{\tilde\lambda}}\tilde\omega_{\rm i}
+2\tilde\Omega_0(2\tilde\Omega_0+\tilde S)
\right.
\ed
\be
\left.
+{{2\pi\kappa_0}\over{\tilde\lambda\tilde H}}(1-\tilde\Omega_0^2)\right]
\tilde\omega_{\rm r}
+{\frac {4\pi^2\tilde\eta}{\tilde\lambda^2\tilde H}}
(1-\tilde\Omega^2_0)=0
\ee
and
\bd
\tilde\omega_{\rm i}^3
-{\frac {2\pi\tilde\eta}{\tilde\lambda}}
\tilde\omega_{\rm i}^2
-\left[ 3\tilde\omega_{\rm r}^2
-2\tilde\Omega_0(2\tilde\Omega_0+\tilde S)\right.
\ed
\be
\left.
-{{2\pi\kappa_0}\over{\tilde\lambda\tilde H}}(1-\tilde\Omega_0^2)
\right]\tilde\omega_{\rm i}
+{\frac {2\pi\tilde\eta}{\tilde\lambda}}
\tilde\omega_{\rm r}^2
-{\frac {4\pi\tilde\eta\tilde \Omega_0  }{\tilde\lambda}}
(2\tilde\Omega_0+\tilde S)=0.
\ee
Let $\omega_{\rm i}=0$ in Eqs. (45) and (46), we have 
\be
\tilde\omega_{\rm r}=\pm
[2\tilde\Omega_0(2\tilde\Omega_0+\tilde S)]^{1/2},
\ee
and
\be
\tilde\eta={\frac {\kappa_0\tilde\lambda}{2\pi}}
\tilde \omega_{\rm r}.
\ee
Since $\eta>0$, $\tilde\omega_{\rm r}/\tilde\lambda$ has the same sign as 
$\kappa_0$. Since the field strength will normally (but not necessarily)
decrease outward in the disk, the inclination of the field lines will be away 
from the axis, i.e. $\kappa>0$. In the absence of diffusion, a transition
from stability to instability, if it exists at all, then is possible only for 
an {\it inward traveling} wave 
mode, i.e. $\tilde\omega_{\rm r}/\tilde\lambda>0$. 

Further progress can be made if we assume that the diffusivity is small, and
by looking at conditions near marginal stability. Treating $\tilde\eta$ it 
as a small quantity of the same order as the other small quantities, i.e.
\bd \tilde\eta\sim\tilde\omega_{\rm i}\sim 1/\tilde\lambda, \ed
the growth rate is then given by
\be
\tilde\omega_{\rm i}\simeq 
{\frac  {A_1} {\tilde\lambda \tilde H(3\tilde\omega_{\rm r}^2 -\omega_0^2)}},
\ee
where 
\be
\omega_0^2=2\tilde\Omega_0(2\tilde\Omega_0+\tilde S)
+2\pi\kappa_0(1-\tilde\Omega_0^2)/\tilde\lambda\tilde H,
\ee
\be
A_1=2\pi\tilde\eta\tilde H\tilde\omega_{\rm r}^2
-4\pi\tilde\eta\tilde H\tilde\Omega_0
(2\tilde\Omega_0+\tilde S).
\ee

The real parts of the roots of the equation are:
\be
\tilde\omega_{\rm r1}\simeq {\frac
{C_1}{\tilde\lambda^2\tilde\omega_0^2\tilde H}},
\ee
where 
\be
C_1=4\pi^2\tilde\eta (1-\tilde\Omega_0^2),
\ee
and
\be
\tilde\omega_{\rm r2,3}\simeq \pm\tilde\omega_0
-{\frac {C_1}{2\tilde\lambda^2\tilde H\tilde\omega_0^2}}.
\ee
The imaginary part of $\tilde\omega$ for these roots is
\be
\tilde\omega_{\rm i1} \simeq {\frac 
{4\pi\tilde\eta\tilde\Omega_0(2\tilde\Omega_0+\tilde S)}
{\tilde\lambda\tilde\omega_0^2}}
\ee
and
\be
\tilde\omega_{\rm i2,3}=
{\frac {\kappa_0 C_1}{2\tilde\lambda^2\tilde H\tilde\omega_0^2}}.
\ee

The modes 2, 3 have the same growth rate in the limit taken here, 
where only the first order in $\tilde \eta$ is kept.
From Eqs. (55) and (56), we find that all three modes are stable
in low-$\tilde\eta$ limit. It implies that the neutral wave and 
the outward traveling wave are always stable. 

The stability condition 
for the inward traveling wave mode is
\be
\tilde\eta<{\frac {\kappa_0\tilde\lambda}{2\pi}}
[2\tilde\Omega_0(2\tilde\Omega_0+\tilde S)]^{1/2},
\ee
Since the inclination $\kappa$ is of order unity, and $\tilde\eta, 
1/\tilde\lambda$ are small numbers, the stability condition is satisfied. 
Not surprisingly, the disk is stable in the absence of a wind torque.

\subsubsection{$\tilde\eta= 0$, $\tilde T_{\rm m0}\ne 0$}

We now include the magnetic torque. In this section we treat it 
as a small quantity of same order as the other small quantities in the 
problem. In the absence of the magnetic diffusion ($\tilde \eta=0$),  
Eqs. (43) and (44) reduce to
\bd
\tilde\omega_{\rm r}^3
-\left[3\tilde\omega_{\rm i}^2
-{\frac {2\tilde\omega_{\rm i}p\tilde T_{\rm m0}}{\tilde\Omega_0}}
+2\tilde\Omega_0(2\tilde\Omega_0+\tilde S)  \right.
\ed
\be
\left. +{{2\pi\kappa_0}\over{\tilde\lambda\tilde H}}(1-\tilde\Omega_0^2)\right]
\tilde\omega_{\rm r}
+{\frac {4\pi q\tilde\Omega_0} {\tilde\lambda\tilde H}}\tilde T_{\rm m0}
-{\frac {4\pi \tilde\Omega_0} {\tilde\lambda\tilde H}}\tilde T_{\rm m0}=0
\ee
and
\bd
\tilde\omega_{\rm i}^3
-{\frac {p\tilde T_{\rm m0}}{\tilde\Omega_0}}
\tilde\omega_{\rm i}^2
-\left[ 3\tilde\omega_{\rm r}^2
-2\tilde\Omega_0(2\tilde\Omega_0+\tilde S)
-{{2\pi\kappa_0}\over{\tilde\lambda\tilde H}}(1-\tilde\Omega_0^2)
\right]\tilde\omega_{\rm i}
\ed
\be
+{\frac {p\tilde T_{\rm m0}}{\tilde\Omega_0}}
\tilde\omega_{\rm r}^2
-{\frac {4\pi q\kappa_0\tilde\Omega_0}{\tilde\lambda\tilde H}}\tilde T_{\rm m0}
-{\frac {2\pi p \kappa_0}{\tilde \lambda\tilde H\tilde\Omega_0}}
(1-\tilde\Omega_0^2)\tilde T_{\rm m0}=0.
\ee

We analyze the situation close to marginal stability in the case 
of low magnetic torque ($\tilde T_{\rm m0}\rightarrow 0$). The growth 
rate is given by
\be
\tilde\omega_{\rm i} \simeq 
{\frac {A_2}{\tilde\lambda\tilde H\tilde\Omega_0(3\tilde\omega_{\rm r}^2
-\tilde\omega_0^2)}},
\ee
where
\be
A_2=\tilde\lambda \tilde H p\tilde T_{\rm m0}\tilde\omega_{r}^2
-4\pi q\kappa_0\tilde\Omega_0^2\tilde T_{\rm m0}
-2\pi p\kappa_0(1-\tilde\Omega_0^2)\tilde T_{\rm m0}.
\ee

The real parts of the roots of the equation are:
\be
\tilde\omega_{\rm r1}\simeq {\frac
{C_2}{\tilde\lambda\tilde H\tilde\omega_0^2}},
\ee
where 
\be
C_2=4\pi q\tilde\Omega_0\tilde T_{\rm m0} 
-4\pi\tilde\Omega_0\tilde T_{\rm m0} 
\ee
and
\be
\tilde\omega_{\rm r2,3}\simeq \pm\tilde\omega_0
-{\frac {C_2}{2\tilde\lambda\tilde H\tilde\omega_0^2}}.
\ee
The imaginary part of $\tilde\omega$ for these roots is
\be
\tilde\omega_{\rm i1} = [{4\pi q\kappa_0\tilde\Omega_0^2
+2\pi p\kappa_0 (1-\tilde\Omega_0^2)]\frac{\tilde T_{\rm m0}}
{\tilde\lambda\tilde H\tilde\Omega_0\tilde\omega_0^2}}\label{om1}
\ee
and
\be
\tilde\omega_{\rm i2,3}=
[p\tilde\lambda\tilde H(2\tilde\Omega_0+\tilde S)-2\pi q \kappa_0\tilde\Omega_0]
\frac{\tilde T_{\rm m0}}{\tilde\lambda\tilde H\tilde\omega_0^2}.\label{om23}
\ee

The modes 2, 3 have the same growth rate in the limit taken here, 
where only the first order in $1/\tilde \lambda$ and $\tilde T_{\rm m0}$ 
is kept.

\subsection{Interpretation of the analytic results}

The growth rates of the three modes (65, 66) are all 
proportional to the magnetic torque. Hence they all qualify as instabilities
caused by the coupling between the disk and the outflow, if their $\omega_{\rm i}$ are 
negative.  The first mode, $\omega_1$ is neutral, in the absence of diffusion and magnetic 
torques. If only magnetic diffusion is included (section 5.2.1), it is a damped mode. 
Thus, low frequency perturbations are stabilized by magnetic diffusion, in the
absence of a magnetic torque. When a magnetic torque is present (section 5.2.2)
the sign of the growth rate is determined [see eq (65)] by the quantity $2q\tilde\Omega^2+ p(1-\tilde\Omega^2)$. 
For low field strength, the second term is small, and 
instability is determined by the sign of $q$, the rate of change of the magnetic
torque with inclination of the field line. As Fig. 1 shows, the magnetic torque 
increases with decreasing inclination (field lines bent further away from the vertical).
This reflects the fact that the centrifugal effect is stronger on field lines that are
bent further away from the vertical, increasing the mass flux and magnetic torque.
Thus $q$ is negative in the cases of interest, and the growth rate $-\omega_{\rm i}$
positive.  The term involving $p$, the dependence of the wind torque on rotation rate,
is stabilizing since $p>0$ (cf Fig.1). In dimensional terms, the approximate condition 
for instability  in the absence of magnetic diffusion and in the limit of a weak magnetic 
wind torque is thus:
\be 
{-\partial T_{\rm m}\over\partial\kappa}> -
{g_{\rm m}\over\kappa}{\partial T_{\rm m}\over\partial g_{\rm m}},
\ee
where $g_{\rm m} = (\Omega_{\rm K}^2-\Omega^2)r_0 = B_r^{\rm S}B_z/(2\pi\Sigma)$ 
is the acceleration due to magnetic support of the disk against gravity.
(the minus signs are included because the derivatives are negative in the cases
of interest).

This mode evidently represents the sought instability, associated with the
dependence of magnetic torque on field line inclination. The result shows, however,
that the instability is not governed only by the dependence of the torque on
inclination. If the magnetic field is strong enough to change the rotation rate by 
providing support against gravity, the second term in (65) is important, and its
effect stabilizing. This represents the fact that a lower rotation rate increases the
height of the potential barrier that the wind has to overcome (cf discussion in
section 3), which decreases the magnetic torque (Ogilvie and Livio, 2001). 

We briefly discuss the two oscillatory modes.  Since $p>0$ and $q<0$ for the 
cases of interest (Fig.1), eq. eq. (66) shows that these modes are stable. 
In the limit of $\tilde\eta=0$ and $\tilde T_{\rm m0}=0$, 
the two modes have the same frequency (cf. Eqs. (54) \& (64))

\be
\tilde\omega_{\rm r2,3}\simeq \pm
\left [2\tilde\Omega_0(2\tilde\Omega_0+\tilde S)
+(1-\tilde\Omega_0^2){{2\pi\kappa_0}\over{\tilde\lambda}}
\right]^{1/2}.
\ee
$[2\tilde\Omega_0(2\tilde\Omega_0+\tilde S)]^{1/2}$ is the 
epicyclic frequency associated with the stable angular momentum
gradient. The second term represents the restoring force due to the
change of inclination of the field lines by the perturbations 
(note that $1-\tilde\Omega_0^2\propto B_r B_z$). On its own, this magnetic 
restoring force would yield a wave with frequency $\omega\propto 
\lambda^{-1/2}$, like a surface wave. This is the result of the
long-range nature of the magnetic perturbations of the potential 
field outside the disk. These magnetic waves have been analyzed before 
(Spruit \& Taam 1990; Tagger et al.  1990).


\section{Numerical Results}

We obtain the growth rate of the instabilities by solving Eq. (40)
numerically. For the shear rate $\tilde S$ we take the Keplerian value 
$S=-3/2$.

The problem is described by the dimensionless 
disk scale-height $\tilde H$, the angular velocity of accretion flow 
$\tilde \Omega_0$ which is a measure of the magnetic field strength, 
the inclination of magnetic field $\kappa_0=B_z/B_r$ at the disk 
surface, and the dimensionless magnetic diffusivity $\tilde\eta$.  
The inclination turns out to influence the stability mainly through its
effect on the wind torque, which depends sensitively on it (Figs.1, 2).
The second main parameter determining the stability is the strength
of the magnetic field, which we have expressed throughout this paper  
by its effect on the rotation rate (relative to the Keplerian rate)
and the angular velocity of accretion flow, $\tilde\Omega_0= 
\Omega/\Omega_{\rm K}$. With increasing field strength, the
magnetic stresses start contributing to the support of the disk 
against gravity, decreasing the rotation rate. 

We take the self-similar index $\alpha$ of the 
magnetic field shape to be $4$ in the following.
In Figs. 1 and 2, the magnetic torque as functions 
of $\kappa_0$ and $\tilde\Omega_0$ is plotted for different 
values of the dimensionless disk scale-height $\tilde H$. 
For small disk scale-height, i.e., if the 
temperature of gas in the accretion disk is low, the magnetic torque is
significant only if the angular velocity is close to Keplerian velocity
and the magnetic field inclination angle to the disk surface is low 
(low $\kappa_0$, see Fig. 2). 
The numerically determined stability boundaries for the neutral  
wave mode are plotted in Figs. 3 and 4. It is seen that instability 
happens in basically two region of parameter space. One is at high
wind torque (right side of the diagrams); the instability then happens 
roughly independent of the strength of the magnetic field. At low
wind torques, instability is possible only when the rotation rate is close
enough to Keplerian.

Examples of the growth rate of the neutral wave mode 
for different parameter values are shown in Figures 5-11.

\begin{figure}
\centerline{\psfig{figure=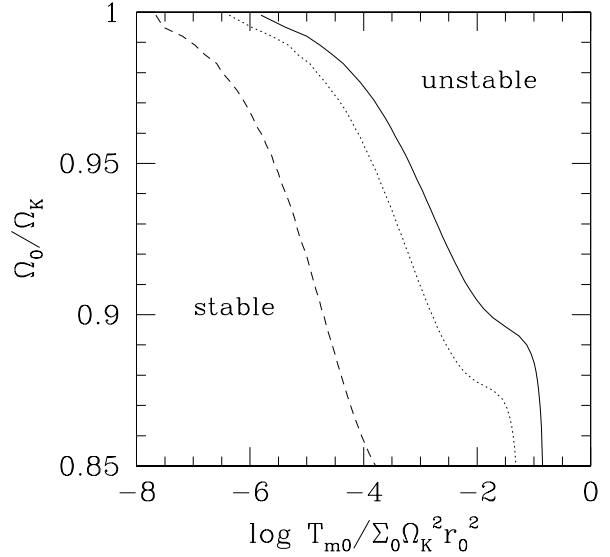,width=1.0\hsize}}
\caption{
The stability boundary of the neutral wave mode, for disk thickness $H/r=0.01$ 
and wavelength $\lambda/H=10$, as a function of wind torque and deviation
from Kepler rotation.
the strength of the magnetic field through the radial equilibrium condition, with high
field strength corresonding to low $\Omega/\Omega_{\rm K}$. The dimensionless diffusivity 
is  $\tilde\eta=0.75$(solid line), 0.25(dotted) and 0.01(dashed). }
\end{figure}

\begin{figure}
\centerline{\psfig{figure=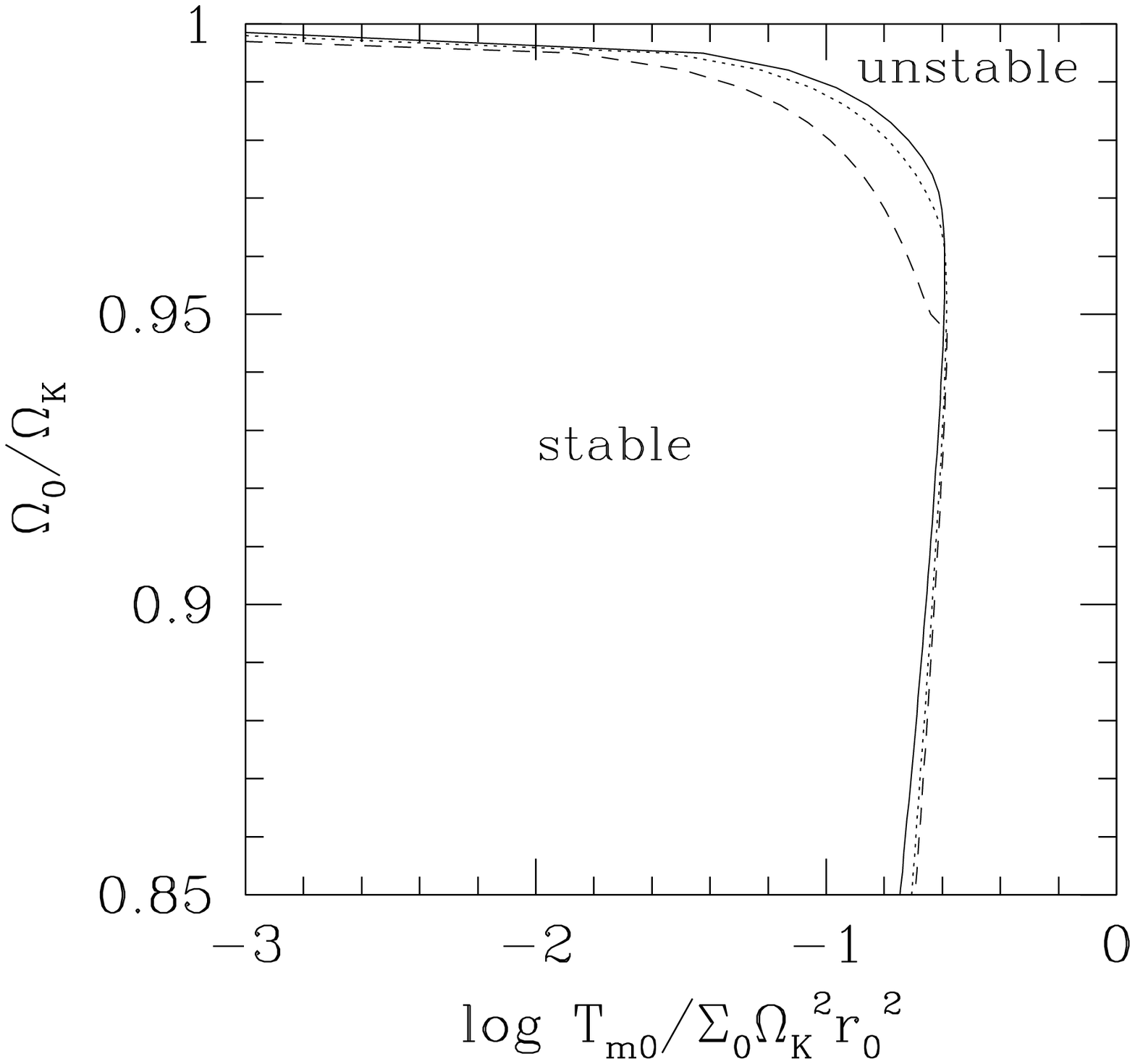,width=1.0\hsize}}
\caption{Same as Fig. 3, but for $H/r=0.001$.  }
\end{figure}

\begin{figure}
\centerline{\psfig{figure=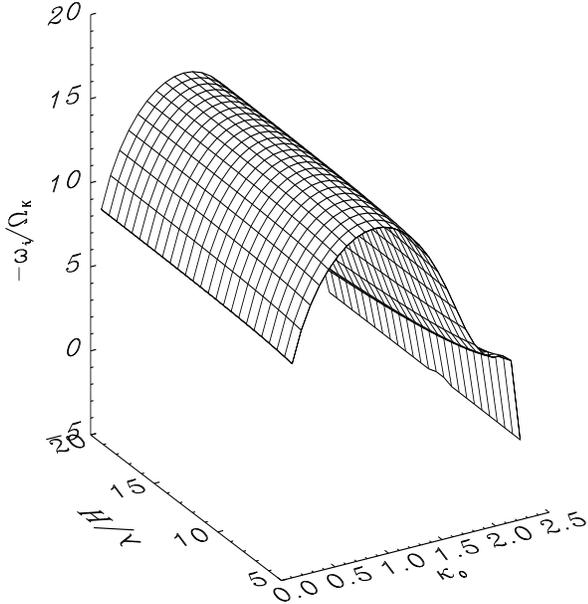,width=10.0cm,height=10.0cm}}
\caption{The growth rate of the instability as functions of 
magnetic field inclination $\kappa_0$ and the dimensionless 
wavelength $\lambda/H$. The dimensionless shearing rate $\tilde S=-3/2$, 
disk thickness  $H/r=0.01$, magnetic diffusivity
$\tilde\eta=0.75$  
and angular velocity $\Omega_0/\Omega_{\rm K}=0.995$ are adopted.
}
\end{figure}

\begin{figure}
\centerline{\psfig{figure=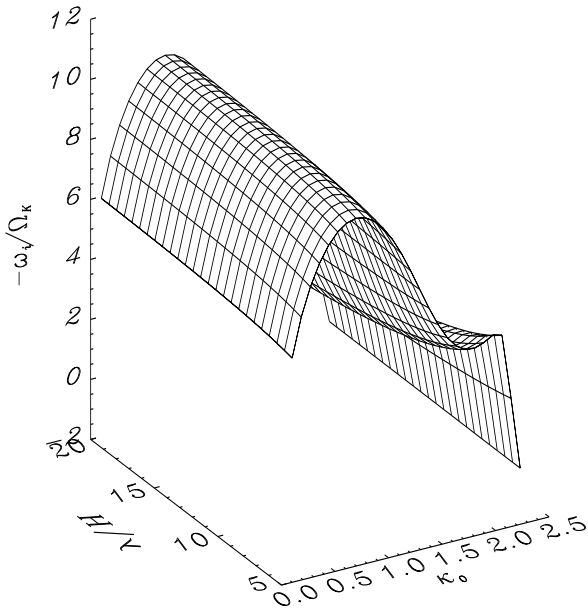,width=10.0cm,height=10.0cm}}
\caption{Same as Fig. 5, but for different value of angular velocity: 
$\Omega_0/\Omega_{\rm K}=0.99$.}
\end{figure}

\begin{figure}
\centerline{\psfig{figure=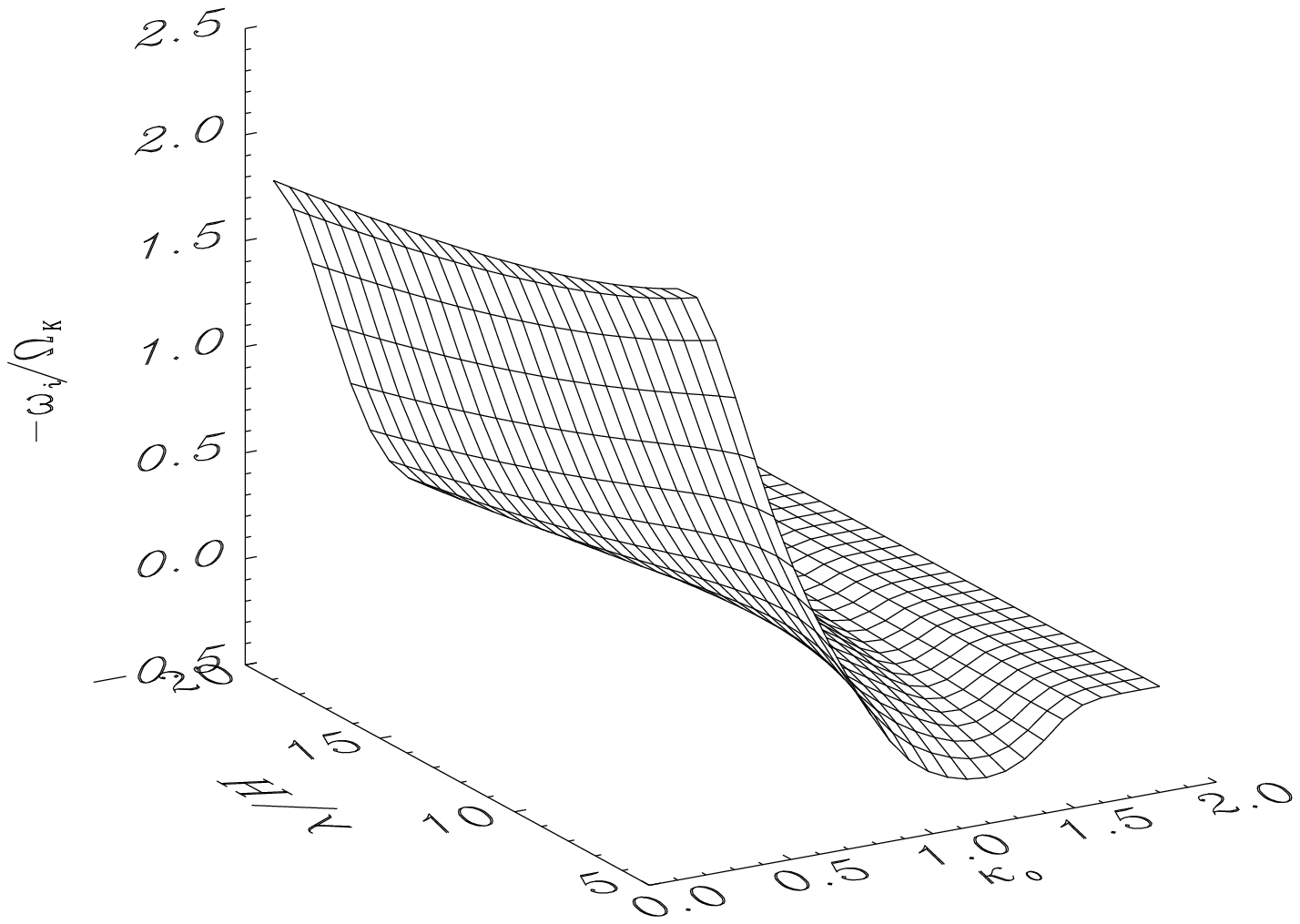,width=10.0cm,height=10.0cm}}
\caption{Same as Fig. 5, but for   
$\Omega_0/\Omega_{\rm K}=0.9$.}
\end{figure}

\begin{figure}
\centerline{\psfig{figure=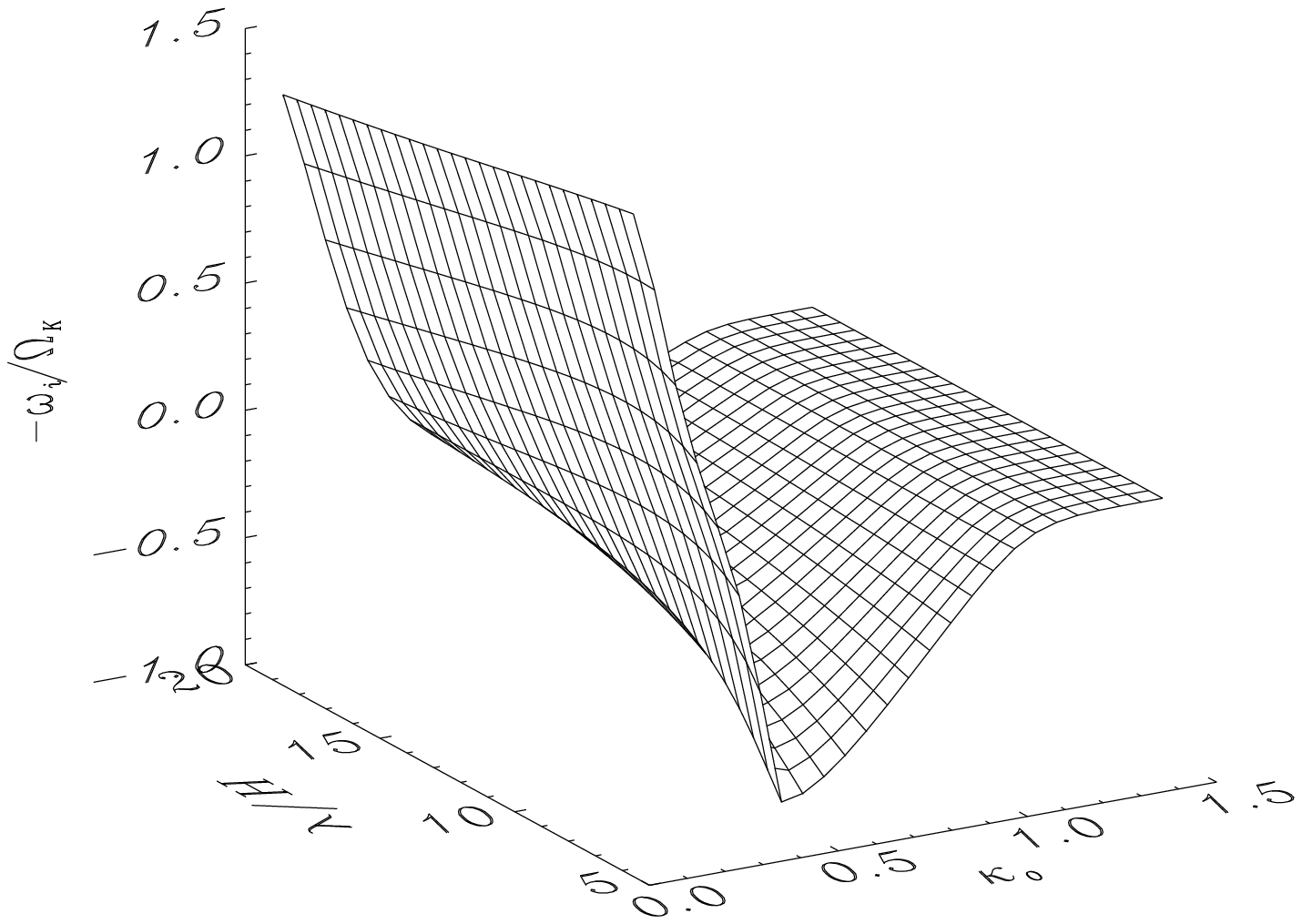,width=10.0cm,height=10.0cm}}
\caption{Same as Fig. 5, but  for 
$\Omega_0/\Omega_{\rm K}=0.85$.}
\end{figure}

\begin{figure}
\centerline{\psfig{figure=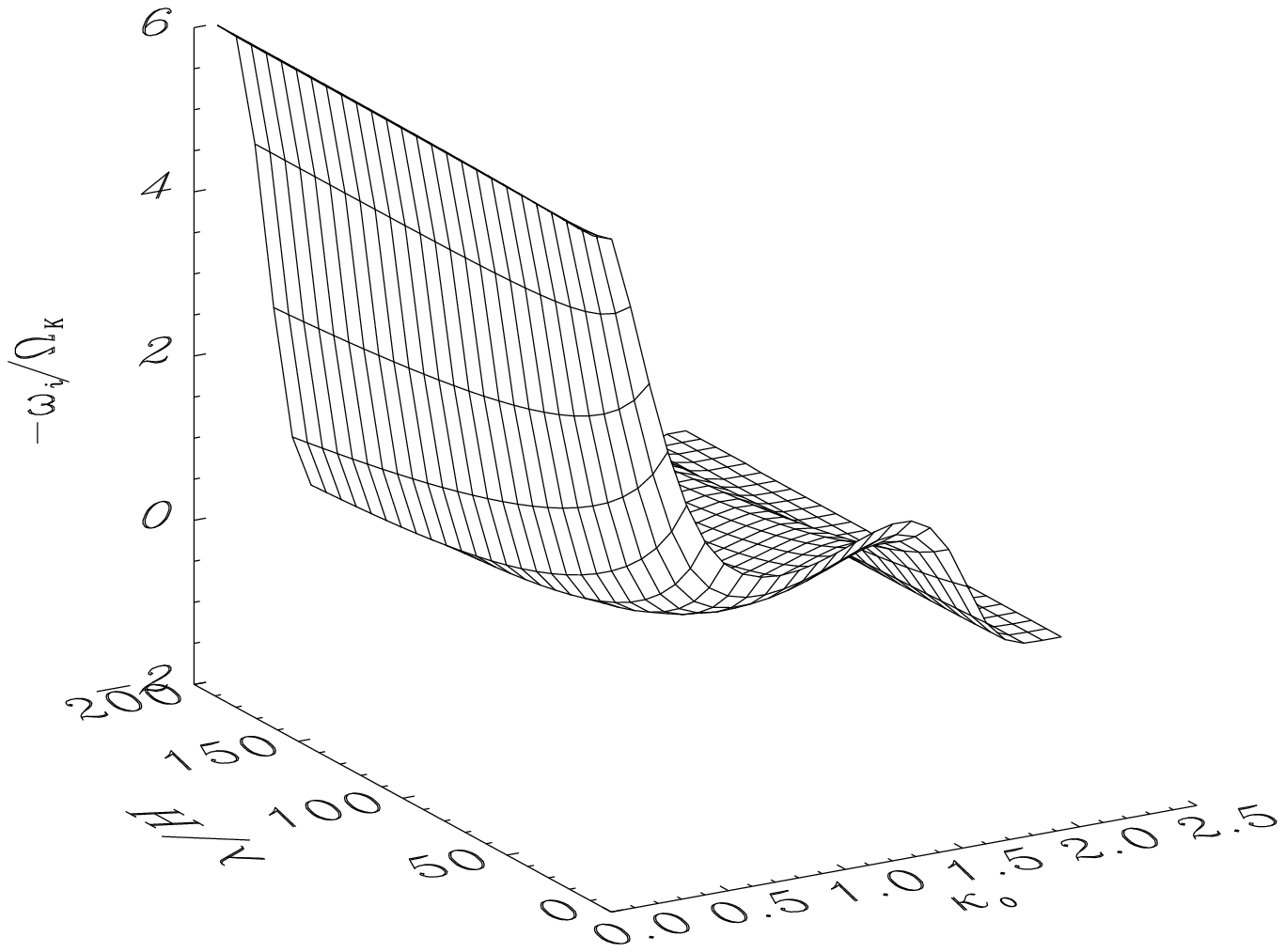,width=10.0cm,height=10.0cm}}
\caption{Same as Fig. 6, but for $H/r=0.001$.}
\end{figure}

\begin{figure}
\centerline{\psfig{figure=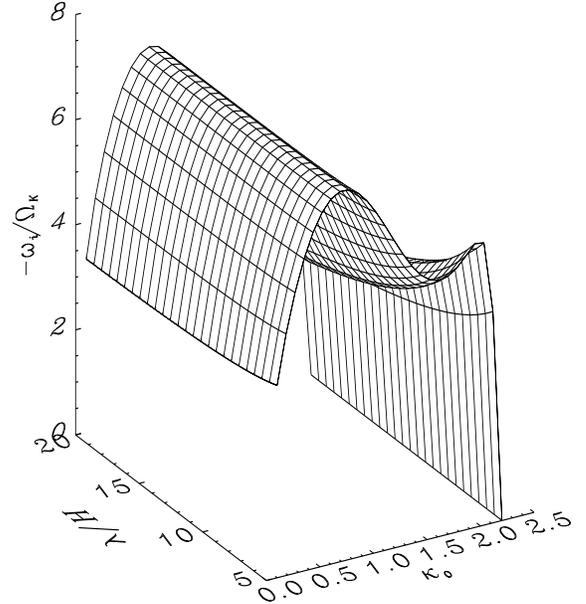,width=10.0cm,height=10.0cm}}
\caption{Same as Fig. 6, but without magnetic diffusion, i.e., 
$\tilde\eta=0$. }
\end{figure}

\begin{figure}
\centerline{\psfig{figure=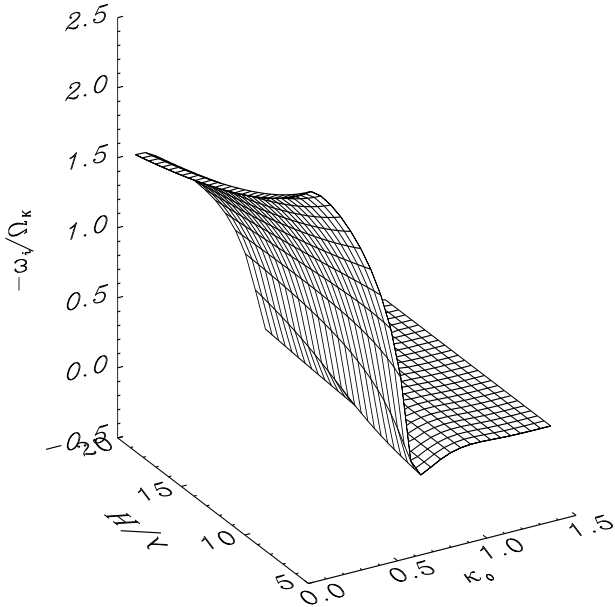,width=10.0cm,height=10.0cm}}
\caption{Same as Fig. 8, but without magnetic diffusion, 
$\tilde\eta=0$. }
\end{figure}

\section{Discussion of the results}

We have presented a linear stability analysis of disks with magnetically 
driven winds, and confirm the instability by the mechanism proposed by LPP. 

In the presence of a strong magnetic field, the disk is 
compressed by the vertical component of gravitational force as 
well as vertical magnetic pressure, and the vertical structure 
of the disk is significantly altered, which affects the position of
sonic point and the wind torque. In this strong field case, the 
angular velocity of accretion flow significantly 
deviates from the Keplerian velocity due to the radial magnetic
force, and the magnetic torque is then negligible for any magnetic 
field inclination angle (see Figs. 1 and 2, see also Ogilvie and Livio 1998). 
We have taken these effects into account in present investigation. 

The system of equations analyzed has four modes. In the absence of a 
magnetic field, two are a neutral displacements. One of these two is 
not specifically related to the magnetic wind torque, and we 
have excluded it from quantitative analysis. The other has a frequency
proportional to the magnetic torque (real and imaginary parts 
of the frequency are of the same order of magnitude),
it can become unstable.


The final two modes represent an inward and
an outward traveling wave. The restoring force in the wave 
is a combination of the coriolis force (epicyclic motion) and the 
magnetic forces.
Both inward and outward traveling waves
are stable in the range of validity of our assumptions.



In Fig. 5, the growth rate of the instability is plotted for angular velocity 
very close to Keplerian velocity ($\tilde\Omega_0=0.995$), which may 
approximate the case considered in LPP. 
The disk becomes more unstable if the angular velocity of the flow is
close to Keplerian (see Figs. 5-8).
For high inclination angles of magnetic field line with respect
to the surface of the disk (large $\kappa_0$), the magnetic torque is
very small. Instability is then suppressed by magnetic diffusion. 

The magnetic torque makes
the disk unstable, while magnetic diffusion has a stabilizing effect, 
suppressing instability at low $\tilde\Omega_0$ and/or torque 
(see Figs. 3 and 4). Comparing Figs. 6 and 9, we see that the disk 
is less unstable in low temperature case ($\tilde H=0.001$) while
the instability occurs for lower values of $\kappa_0$.

To see the effect of magnetic diffusion, Figs. 10 and 
11 show the growth rates for the cases with $\tilde\eta=0$. In the 
absence of magnetic diffusion, the disk is always unstable, though
the growth rates can be very small. 

The physical reason for the instability found here has been described 
in LPP (see discussion in section 2). 
A perturbation increasing $v_r$ causes the poloidal field to be 
bent close to the disk plane, as the field is advected inwards by the 
accreting matter. This tends to increase the mass flow along the 
field. There is also an opposing effect, however. As the poloidal 
magnetic field is bent towards the disk plane, the radial curvature 
force on the disk (opposite to gravitational force increases). Due 
to this increased support against gravity, the rotation rate of the field
line decreases. As a result, the ``potential barrier'' for mass flowing
along the field line is larger, decreasing the mass flux and magnetic
torque. The instability decreases with increasing field strength through 
its effect on the rotation rate, since the instability is driven by the
magnetic torques. We find that the inclination of the magnetic field 
still has the strongest influence on the magnetic torque, and
dominates over the change in the potential barrier. 

Since the instability is caused by the magnetic torque by the
centrifugally driven wind from the disk, the instability disappears 
as the magnetic torque vanishes (Figs. 3 and 4). The disk is stable 
without a magnetic torque, all three modes are then damped by
magnetic diffusion. 
If the magnetic torque is large, the instability time scale is comparable 
with the dynamical time scale of the disk, and the instability is not 
much affected by magnetic diffusion (compare Figs. 6 and 10). 
We find that the instability is insensitive to the perturbation wavelength.

Associated with the perturbations in the disk are changes of the field configuration
above the disk. These affect the acceleration of the wind, and hence
the wind torque. In our analysis, we have taken the response of the wind
torque to these changes to be instantaneous. This is motivated by the fact that
the magnetic torque acts on the disk surface, and changes as soon as the 
azimuthal field component at the surface changes. This is perhaps somewhat
contrary to the impression that one might have based on the steady wind model.
In the steady wind model, the torque is often pictured as `effectively acting' at
the Alfv\'en surface. In reality, the torque is constant along the field line,
hence acts equally at the disk surface. When changes due to motions in the disk
take place, changes in the torque travel up at the Alfv\'en speed and are first 
felt in disk itself. 

The actual propagation of these torque changes has not
been considered here, since its time scale is short, of the order of the
Alfv\'en travel time over the disk thickness. The time scale $\tau_{\rm Ad}$ 
for an Alfv{\'e}n wave to travel over the scale-height is of the order
$\tau_{\rm Ad}\approx c_{\rm s}/V_{\rm A}\Omega^{-1}$, which is short except 
for weak fields. The time scale for Alfv{\'e}n wave traveling from 
the disk surface to the Alfv{\'e}n point in the wind is in fact shorter than 
this, $\sim 1/\Omega_0$. The disk-wind coupling time scale is therefore of 
the order of the disk dynamical time (Wardle \& K\"onigl 1993; K\"onigl \& 
Wardle 1996). 
The present analyses are valid if the instability growth time scale is much 
longer than the disk-wind coupling time scale. This is the case since for 
weak fields, where the coupling time is long, the growth time of the 
instability itself is also long.

In order to check on the effect of finite  Alfv{\'e}n travel times in the 
disk-wind coupling process on the 
instability properties of the disk, we have done a test calculation
with a simplified description of this effect. In this calculation, 
we manually induce a time delay 
of the dynamical time scale of the disk on the wind torque:

\be 
{\frac {\partial T_{\rm m}(t)}{\partial t}}={\frac 
{T_{\rm m0}-T_{\rm m}(t)}{\tau_{\rm delay}}},
\ee
where $T_{\rm m0}$ is given by Eq. (15), $\tau_{\rm delay}$ is the 
assumed disk-wind coupling time scale. We then add this equation to 
the set of disk equations. We find the growth rate of the instability
becomes slightly lower than before, but the instability properties 
have not been altered qualitatively. 

The nonlinear evolution of the instability analyzed gas to be studied by 
numerical simulations. Some such simulations have already been 
reported by Agapitou \& Papaloizou (2001, in preparation).

\acknowledgements{
This work was done in the context of the TMR Network `accretion 
onto black holes' (European Commission grant ERBFMRX-CT98-0195).  
XC is supported by CAS-MPG exchange program. He thanks the support from  
NSFC, the NKBRSF (No. G1999075403), and Pandeng Project.}    

{}

\end{document}